\documentclass{article}
\usepackage{spconf, graphicx}
\usepackage{amsmath}
\usepackage{amssymb}
\newtheorem{remark}{Remark}

\title{Interactive Model Fusion-Based GM-PHD Filter}
\twoauthors
  {Jiacheng He, Shan Zhong, Bei Peng\sthanks{Thanks to the NSFC 51975107, Sichuan Provincial Science and Technology Major Program 2022ZDZX0039, and No. 2019 ZDZX0020 for financial support of the work.}}
	{School of Mechanical and\\ Electrical Engineering\\
	University of Electronic Science and\\
       Technology of China\\
	Chengdu P.R. China 611731}
  {Gang Wang, Qizhen Wang\sthanks{Thanks to Sichuan Science and Technology Program No. 2022YFG0343 for financial support of the work.}}
	{School of Information and\\ Communication Engineering\\
	University of Electronic Science and\\ Technology of China\\
	Chengdu P.R. China 611731}

\begin{document}
%
\maketitle
\begin{abstract}
In multi-target tracking (MTT), non-Gaussian measurement noise from sensors can diminish the performance of the Gaussian-assumed Gaussian mixture probability hypothesis density (GM-PHD) filter. In this paper, an approach that transforms the MTT problem under non-Gaussian conditions into an MTT problem under Gaussian conditions is developed. Specifically, measurement noise with a non-Gaussian distribution is modeled as a weighted sum of different Gaussian distributions. Subsequently, the GM-PHD filter is applied to compute the multi-target states under these distinct Gaussian distributions. Finally, an interactive multi-model framework is employed to fuse the diverse multi-target state information into a unified synthesis. The effectiveness of the proposed approach is validated through the simulation results.

\end{abstract}
\section{Introduction}
\label{sec:intro}
MTT \cite{9746344,10096575} plays a pivotal role in various scenarios, including ground traffic control, remote sensing, computer vision, and more. The classical PHD is an elegant approach for handling MTT, moreover, the GM-PHD \cite{1710358} effectively addresses the complex multiple integration challenges within the PHD recursive process. The GM-PHD method has significantly propelled the realization of large-scale target tracking. However, the GM-PHD filter is better suited for scenarios with Gaussian noise only, and its performance is inevitably influenced by non-Gaussian noise \cite{fan2022background,he2023generalized,he2023maximum}. 
This paper primarily focuses on the MTT problem with non-Gaussian noise.

To mitigate the effect of non-Gaussian sensor noise on the performance of the PHD filter, the sequential Monte Carlo-based PHD (SMC-PHD) algorithm \cite{vo2005sequential} was developed. The SMC-PHD algorithm enhances accuracy at the cost of increased computational complexity, which falls short of meeting the real-time requirements in practical applications. To enhance the efficiency of the SMC-PHD-type algorithms, an auxiliary SMC-PHD (ASMC-PHD) filter \cite{5545199} was developed. Similarly, the Spline PHD (SPHD) filter \cite{sithiravel2013spline} was devised, leveraging the properties of B-splines to approximate arbitrary probability density functions. Constrained by the computational complexity, the BPHD algorithm is well-suited for tracking a limited number of high-value targets. 

To develop computationally efficient MTT algorithms under non-Gaussian conditions, the Student's $t$ mixture PHD filter was developed for a certain type of non-Gaussian noise with heavy-tailed distribution, modeling the noise using Student's $t$ distribution. The idea was also utilized in the development of the Student’s $t$ mixture cardinality-balanced multi-target multi-Bernoulli filter \cite{wang2018student} and the labeled multi-Bernoulli filter \cite{dong2018student}. The MTT algorithms based on the Student's $t$ distribution effectively mitigate the adverse impacts with lower computational cost. However, when dealing with non-Gaussian noise characterized by a more general distribution, it inevitably affects the performance of the algorithms relying on the specific Student's $t$ distribution.

The core strategy of the proposed methodology involves the conversion of the MTT challenge under non-Gaussian noise into an equivalent multi-target tracking challenge under Gaussian noise. To elaborate further, measurement noise with a non-Gaussian distribution is characterized by a set of Gaussian distributions with different means, variances, and weights. These Gaussian distributions are individually incorporated into the conventional GM-PHD algorithm, generating disparate multi-target state information. Ultimately, employing an interactive multi-model framework, the diverse state information is amalgamated into a unified synthesis. Based on the above idea, the method we propose is called the GM-PHD filter based on interactive model fusion (IMF-GM-PHD).

\section{Problem formulation}
\label{sec:format}
For a class of MTT problems, the state and measurement variables of multi-target are described as random finite sets (RFS):
\begin{align}
\left\{ \begin{gathered}
  {{\boldsymbol{X}}_k} \triangleq \left\{ {{{\boldsymbol{x}}_{k,1}}, \cdots ,{{\boldsymbol{x}}_{k,{n_x}}}} \right\} \in \Re \left( {\boldsymbol{X}} \right), \hfill \\
  {{\boldsymbol{Z}}_k} \triangleq \left\{ {{{\boldsymbol{z}}_{k,1}}, \cdots ,{{\boldsymbol{z}}_{k,{m_k}}}} \right\} \in \Re \left( {\boldsymbol{Z}} \right). \hfill \\ 
\end{gathered}  \right. 
\end{align}
Here ${n_x}$ and ${m_x}$ stand for the number of targets and measurement data, and ${{\boldsymbol{z}}_{k,{m_k}}} \in {\mathbb{R}^{m \times 1}}$; $\Re \left( {\boldsymbol{X}} \right)$ and $\Re \left( {\boldsymbol{Z}} \right)$ denote the state and measurement space of multi-target. In areas detected by sensors, targets are typically subject to various events, including target emergence, target derivation from existing ones, and target disappearance. Therefore, at the $k$th instant, the states of multi-target are represented as
\begin{align}
{{\boldsymbol{X}}_k} = \left[ { \cup {{\boldsymbol{S}}_{k|k - 1}}\left( {\boldsymbol{\varsigma }} \right)} \right] \cup \left[ { \cup {{\boldsymbol{B}}_{k|k - 1}}\left( {\boldsymbol{\varsigma }} \right)} \right] \cup {{\boldsymbol{\Gamma }}_k},
\end{align}
where ${{{\boldsymbol{S}}_{k|k - 1}}\left( {\boldsymbol{\varsigma }} \right)}$ represents the RFS representing the existing targets at the next time step, which includes either target survival or target disappearance. ${p_{S,k}}\left( {{{\boldsymbol{x}}_{k - 1}}} \right)$ represents the probability of target survival, while the probability of target disappearance is represented as $1 - {p_{S,k}}\left( {{{\boldsymbol{x}}_{k - 1}}} \right)$. ${{{\boldsymbol{B}}_{k|k - 1}}\left( {\boldsymbol{\varsigma }} \right)}$ stands for the RFS of targets spawned from state ${\boldsymbol{\varsigma }}$, and ${{\boldsymbol{\Gamma }}_k}$ represents the RFS of target that birth naturally. Furthermore, the measurement information acquired by sensors is modeled as:
\begin{align}
{{\boldsymbol{Z}}_k} = \left[ {\mathop  \cup \limits_{{\boldsymbol{x}} \in {{\boldsymbol{X}}_k}} {{\boldsymbol{\Theta }}_k}\left( {{{\boldsymbol{x}}_k}} \right)} \right] \cup {{\boldsymbol{K}}_k}.
\end{align}
Here ${{{\boldsymbol{\Theta }}_k}\left( {\boldsymbol{x}} \right)}$ stands for the RFS generated from ${{{\boldsymbol{x}}_k}}$. The probability of ${{{\boldsymbol{x}}_k}}$ being detected is denoted as ${p_{D,k}}\left( {{{\boldsymbol{x}}_k}} \right)$, while the probability of not being detected is $1 - {p_{D,k}}\left( {{{\boldsymbol{x}}_k}} \right)$. ${{\boldsymbol{K}}_k}$ stands for the RFS of clutter. 

The posterior estimation can be obtained through optimal Bayesian filtering in \cite{1710358}, but this approach is often infeasible in practice. The PHD filter provides a feasible approach for MTT by recursively propagating the first-order statistical characteristics of the posterior density. The PHD iteration is
\begin{align}
\begin{gathered}
  {v_{k|k - 1}}\left( {\boldsymbol{x}} \right) = {\gamma _k}\left( {\boldsymbol{x}} \right) +  \hfill \\
  \int {\left[ {{p_{S,k}}{{\boldsymbol{f}}_{k|k - 1}}\left( {{\boldsymbol{x}}|{\boldsymbol{\varsigma }}} \right) + {\beta _{k|k - 1}}\left( {{\boldsymbol{x}}|{\boldsymbol{\varsigma }}} \right)} \right]{v_{k - 1}}\left( {\boldsymbol{x}} \right)d{\boldsymbol{\varsigma }}}  \hfill \\ 
\end{gathered} 
\end{align}
and
\begin{align}
\begin{gathered}
  {v_k}\left( {\boldsymbol{x}} \right) = \left[ {1 - {p_{D,k}}\left( {\boldsymbol{x}} \right)} \right]{v_{k|k - 1}}\left( {\boldsymbol{x}} \right) +  \hfill \\
  \sum\limits_{{{\boldsymbol{z}}_k} \in {{\boldsymbol{Z}}_k}} {\frac{{{p_{D,k}}\left( {\boldsymbol{x}} \right){{\boldsymbol{g}}_k}\left( {{\boldsymbol{z}}|{\boldsymbol{x}}} \right){v_{k|k - 1}}\left( {\boldsymbol{x}} \right)}}{{{\kappa _k}\left( {{{\boldsymbol{z}}_k}} \right) + \int {{p_{D,k}}\left( {\boldsymbol{\varsigma }} \right){{\boldsymbol{g}}_k}\left( {{\boldsymbol{z}}|{\boldsymbol{\varsigma }}} \right){v_{k|k - 1}}\left( {\boldsymbol{\xi }} \right)d\left( {\boldsymbol{\varsigma }} \right)} }}} , \hfill \\ 
\end{gathered}  
\end{align}
where ${v_{k|k - 1}}\left( {\boldsymbol{x}} \right)$ and ${v_k}\left( {\boldsymbol{x}} \right)$ correspond to the predictive intensity and updated intensity of the multi-target posterior density, respectively. ${\beta _{k|k - 1}}\left( { \cdot |{\boldsymbol{\varsigma }}} \right)$ and ${\gamma _k}\left( {\boldsymbol{x}} \right)$ represent the intensities of the RFS ${B_{k|k - 1}}\left( {\boldsymbol{\varsigma }} \right)$ and ${{\boldsymbol{\Gamma }}_k}$. ${\kappa _k}\left(  \cdot  \right)$ represents the intensity of the clutter RFS. ${{\boldsymbol{f}}_{k|k - 1}}\left( { \cdot | \cdot } \right)$ stands for the state transition function and ${{\boldsymbol{g}}_k}\left( { \cdot | \cdot } \right)$ is the probability density of the receiving the measurement, and it is commonly assumed that they adhere to the following Gaussian model:
\begin{align}
\left\{ \begin{gathered}
  {{\boldsymbol{f}}_{k|k - 1}}\left( {{\boldsymbol{x}}|{\boldsymbol{\varsigma }}} \right) = \mathcal{N}\left( {{\boldsymbol{x}};{{\boldsymbol{F}}_{k - 1}}{\boldsymbol{\varsigma }},{{\boldsymbol{Q}}_{k - 1}}} \right), \hfill \\
  {{\boldsymbol{g}}_k}\left( {{\boldsymbol{z}}|{\boldsymbol{x}}} \right) = \mathcal{N}\left( {{\boldsymbol{z}};{{\boldsymbol{H}}_k}{\boldsymbol{x}},{{\boldsymbol{R}}_k}} \right), \hfill \\ 
\end{gathered}  \right.
\end{align}
where $\mathcal{N}\left( { \cdot ;{\boldsymbol{m}},{\boldsymbol{P}}} \right)$ represents Gaussian density, and its mean and covariance are ${\boldsymbol{m}}$ and ${\boldsymbol{P}}$. ${{{\boldsymbol{F}}_{k - 1}}}$ denotes the state transition matrix, ${{{\boldsymbol{Q}}_{k - 1}}}$ denotes the process noise covariance, ${{{\boldsymbol{H}}_k}}$ represents the measurement matrix, and ${{{\boldsymbol{R}}_k}}$ stands for the measurement noise covariance matrix. Nonetheless, owing to impulsive disturbances, outliers, or other factors, the distributions of ${{\boldsymbol{g}}_k}\left( { \cdot | \cdot } \right)$ are typically no longer Gaussian. Such non-Gaussian distributions will degrade the effectiveness of the PHD filter. 
In our study, a new method is developed to address the effects of non-Gaussian noise.

\section{The proposed IMF-GM-PHD filter}
The initial step in our approach involves decomposing the non-Gaussian distribution into multiple Gaussian distributions, where the combination of these Gaussian components can serve as an approximate substitute for the original distribution. In our study, we employ the Gaussian mixture model (GMM) \cite{bishop2006pattern} to carry out the decomposition of the measurement noise distributions. Consequently, the probability density function (PDF) of the original measurement noise distribution is approximated as 
\label{sec:pagestyle}
\begin{align}
p\left( {{{\boldsymbol{r}}_k}} \right) = \sum\limits_{l = 1}^L {{\delta ^l}\mathcal{N}\left( {{{\boldsymbol{r}}_k};{\boldsymbol{\mu }}_k^l,{\boldsymbol{R}}_k^l} \right)} ,\sum\limits_{l = 1}^L {{\delta ^l}}  = 1,
\end{align}
with
\begin{align}
\mathcal{N}\left( {{{\boldsymbol{r}}_k};{\boldsymbol{\mu }}_k^l,{\boldsymbol{R}}_k^l} \right) = \frac{{\exp \left\{ { - \tfrac{1}{2}{{\left( {{\boldsymbol{e}}_{ru}^l} \right)}^T}{{\left( {{\boldsymbol{R}}_k^l} \right)}^{ - 1}}{\boldsymbol{e}}_{ru}^l} \right\}}}{{{{\left( {2\pi } \right)}^{m/2}}{{\left| {{\boldsymbol{R}}_k^l} \right|}^{1/2}}}},
\end{align}
where ${\boldsymbol{e}}_{ru}^l = {{\boldsymbol{r}}_k} - {\boldsymbol{\mu }}_k^l$, $l$ is the number of the Gaussian component, and ${{{\boldsymbol{r}}_k}}$ denotes the measurement noise; ${{\delta ^l}}$, ${{\boldsymbol{\mu }}_k^l}$, and ${{\boldsymbol{R}}_k^l}$ stand for the proportion, mean, covariance of the $l$th Gaussian distribution; $\left| {\boldsymbol{A}} \right|$ represents the determinant of matrix ${\boldsymbol{A}}$ and ${\left(  \cdot  \right)^T}$ is the transpose of a matrix.

Utilizing GMM to obtain the proportions ${{\delta ^l}}$ of each Gaussian component, thus obtaining the initial Markov transition probability matrix in interactive multi-model framework as follows
\begin{align}
{\boldsymbol{\tilde P}} = \left[ {\begin{array}{*{20}{c}}
  {{\delta ^1}}&{{\delta ^2}}& \cdots &{{\delta ^L}} \\ 
  {{\delta ^1}}&{{\delta ^2}}& \cdots &{{\delta ^L}} \\ 
   \vdots & \vdots & \ddots & \vdots  \\ 
  {{\delta ^1}}&{{\delta ^2}}& \cdots &{{\delta ^L}} 
\end{array}} \right].
\end{align}
Furthermore, the element in the $l$th row and $u$th column of matrix ${\boldsymbol{\tilde P}}$ is used to represent the transition probability from model $l$ to model $u$, denoted as ${{{[{\boldsymbol{\tilde P}}]}_{lu}}}$.

The key steps of the IMF-GM-PHD filter algorithm are outlined through the subsequent primary steps. In the IMF-GM-PHD filter, the form of a Gaussian mixture is assumed for the posterior intensity at instant $k-1$
\begin{align}\label{pkj1imkixf}
{v_{k - 1}}\left( {\boldsymbol{x}} \right) = \sum\nolimits_{i = 1}^{{J_{k - 1}}} {w_{k - 1}^i\mathcal{N}\left( {{\boldsymbol{x}};{\boldsymbol{m}}_{k - 1}^i,{\boldsymbol{P}}_{k - 1}^i} \right)} ,
\end{align}
where ${{J_{k - 1}}}$, ${w_{k - 1}^i}$, ${{\boldsymbol{m}}_{k - 1}^i}$, and ${{\boldsymbol{P}}_{k - 1}^i}$ decide the characteristics of the posterior intensity. ${{\boldsymbol{m}}_{k - 1}^i}$ is the $i$th peak of the posterior intensity, and the spread of the birth intensity near ${{\boldsymbol{m}}_{k - 1}^i}$ is determined by ${{\boldsymbol{P}}_{k - 1}^i}$. ${w_{k - 1}^i}$ represents the anticipated count of new targets emerging from ${{\boldsymbol{P}}_{k - 1}^i}$.

\begin{remark}
This article involves two types of Gaussian components. One type is used to describe the distribution of intensities (indexed as 'i'), while the other type represents the distribution of observation noise (indexed as 'l' and 'u').
\end{remark}

1) Parameter initialization
\begin{align}
\left\{ \begin{gathered}
  \vartheta _{k - 1}^{i,l} = {\delta ^l},w_{k - 1}^{i,l} = w_{k - 1}^i, \hfill \\
  {\boldsymbol{m}}_{k - 1}^{i,l} = {\boldsymbol{m}}_{k - 1}^i,{\boldsymbol{P}}_{k - 1}^{i,l} = {\boldsymbol{P}}_{k - 1}^i, \hfill \\ 
\end{gathered}  \right.
\end{align}
where $\vartheta _{k - 1}^{i,l}$ represents the model probability corresponding to the $l$th model of the $i$th Gaussian component in \eqref{pkj1imkixf}.

2) Input Interaction: For the $i$th Gaussian component, calculate the probability from model $l$ to $u$.
\begin{align}
\left\{ \begin{gathered}
  \vartheta _{k - 1}^{i,lu} = {{{{[{\boldsymbol{\tilde P}}]}_{lu}}\vartheta _{k - 1}^{i,l}} \mathord{\left/
 {\vphantom {{{{[{\boldsymbol{\tilde P}}]}_{lu}}\vartheta _{k - 1}^{i,l}} {{{\bar c}_u}}}} \right.
 \kern-\nulldelimiterspace} {{{\bar c}_u}}}, \hfill \\
  {{\bar c}_u} = \sum\nolimits_{u = 1}^L {{{[{\boldsymbol{\tilde P}}]}_{lu}}\vartheta _{k - 1}^{i,l}} . \hfill \\ 
\end{gathered}  \right.
\end{align}
Obtain fused Gaussian component information $\bar w_{k - 1}^{i,u}$, ${\boldsymbol{\bar m}}_{k - 1}^{i,u}$, and ${\boldsymbol{\bar P}}_{k - 1}^{i,u}$ using
\begin{align}
\left\{ \begin{gathered}
  \bar w_{k - 1}^{i,u} = \sum\limits_{l = 1}^L {w_{k - 1}^{i,l}\vartheta _{k - 1}^{i,lu}} ,{\boldsymbol{\bar m}}_{k - 1}^{i,u} = \sum\limits_{l = 1}^L {{\boldsymbol{m}}_{k - 1}^{i,l}\vartheta _{k - 1}^{i,lu},}  \hfill \\
  {\boldsymbol{\bar P}}_{k - 1}^{i,u} = \sum\nolimits_{l = 1}^L {\vartheta _{k - 1}^{i,lu}\left[ {{\boldsymbol{P}}_{k - 1}^{i,l} + {{\boldsymbol{e}}_m}{\boldsymbol{e}}_m^T} \right]} , \hfill \\ 
\end{gathered}  \right.
\end{align}
with ${{\boldsymbol{e}}_m} = {\boldsymbol{m}}_{k - 1}^{i,l} - {\boldsymbol{\bar m}}_{k - 1}^{i,u}$.

3) Perform the GM-PHD filter for the $u$th model. Moreover, the covariance of the observation noise obtained through GMM is different for these PHD filters. The predicted intensity of the $u$th model is given
\begin{align}
v_{k|k - 1}^u\left( {\boldsymbol{x}} \right) = v_{S,k|k - 1}^u\left( {\boldsymbol{x}} \right) + v_{\beta ,k|k - 1}^u\left( {\boldsymbol{x}} \right) + {\gamma _k}\left( {\boldsymbol{x}} \right),
\end{align}
with
\begin{align}
\left\{ \begin{gathered}
  v_{S,k|k - 1}^u\left( {\boldsymbol{x}} \right) =  \hfill \\
  {p_{S,k}}\sum\nolimits_{i = 1}^{{J_{k - 1}}} {\bar w_{k - 1}^{i,u}\mathcal{N}\left( {{\boldsymbol{x}};{\boldsymbol{\bar m}}_{S,k|k - 1}^{i,u},{\boldsymbol{\bar P}}_{S,k|k - 1}^{i,u}} \right)} , \hfill \\
  {\boldsymbol{\bar m}}_{S,k|k - 1}^{i,u} = {{\boldsymbol{F}}_{k - 1}}{\boldsymbol{\bar m}}_{k - 1}^{i,u}, \hfill \\
  {\boldsymbol{\bar P}}_{S,k|k - 1}^{i,u} = {{\boldsymbol{Q}}_{k - 1}} + {{\boldsymbol{F}}_{k - 1}}{\boldsymbol{\bar P}}_{k - 1}^{i,u}{\boldsymbol{F}}_{k - 1}^T, \hfill \\ 
\end{gathered}  \right.
\end{align}
and
\begin{align}
\left\{ \begin{gathered}
  {v_{\beta ,k|k - 1}}\left( {\boldsymbol{x}} \right) =  \hfill \\
  \sum\limits_{i = 1}^{{J_{k - 1}}} {\sum\limits_{\tau  = 1}^{{J_{\beta ,k}}} {\bar w_{k - 1}^{i,u}\bar w_{k - 1}^\tau \mathcal{N}\left( {{\boldsymbol{x}};{\boldsymbol{\bar m}}_{\beta ,k|k - 1}^{i,\tau ,u},{\boldsymbol{\bar P}}_{S,k|k - 1}^{i,\tau ,u}} \right)} } , \hfill \\
  {\boldsymbol{\bar m}}_{\beta ,k|k - 1}^{i,\tau ,u} = {\boldsymbol{F}}_{\beta ,k - 1}^\tau {\boldsymbol{\bar m}}_{k - 1}^{i,u} + d_{\beta ,k - 1}^\tau , \hfill \\
  {\boldsymbol{\bar P}}_{\beta ,k|k - 1}^{i,\tau ,u} = {\boldsymbol{Q}}_{\beta ,k - 1}^\tau  + {\boldsymbol{F}}_{\beta ,k - 1}^\tau {\boldsymbol{\bar P}}_{\beta ,k - 1}^{i,u}{\left( {{\boldsymbol{F}}_{\beta ,k - 1}^\tau } \right)^T}, \hfill \\ 
\end{gathered}  \right.
\end{align}
Suppose that the predicted intensity is in the form of a Gaussian mixture
\begin{align}
v_{k|k - 1}^u\left( {\boldsymbol{x}} \right) = \sum\limits_{i = 1}^{{J_{k|k - 1}}} {w_{k|k - 1}^{i,u}\mathcal{N}\left( {{\boldsymbol{x}};{\boldsymbol{\bar m}}_{k|k - 1}^{i,u},{\boldsymbol{\bar P}}_{k|k - 1}^{i,u}} \right)} ,
\end{align}
and the posterior intensity also takes the form of a Gaussian mixture, as follows:
\begin{align}
v_k^u\left( {\boldsymbol{x}} \right) = \left( {1 - {p_{D,k}}} \right)v_{k|k - 1}^u\left( {\boldsymbol{x}} \right) + \sum\limits_{{\boldsymbol{z}} \in {{\boldsymbol{Z}}_k}} {v_{D,k}^u\left( {{\boldsymbol{x}};{\boldsymbol{z}}} \right)} ,
\end{align}
with
\begin{align}
v_{D,k}^u\left( {{\boldsymbol{x}};{\boldsymbol{z}}} \right) = \sum\limits_{i = 1}^{{J_{k|k - 1}}} {w_k^{i,u}\left( {\boldsymbol{z}} \right)\mathcal{N}\left( {{\boldsymbol{x}};{\boldsymbol{m}}_{k|k}^{i,u},{\boldsymbol{P}}_{k|k}^{i,u}} \right)} ,
\end{align}

\begin{align}
w_k^{i,u}\left( {\boldsymbol{z}} \right) = \frac{{{p_{D,k}}w_{k|k - 1}^{i,u}q_k^{i,u}\left( {\boldsymbol{z}} \right)}}{{{\kappa _k}\left( {\boldsymbol{z}} \right) + {p_{D,k}}\sum\nolimits_{\tau  = 1}^{{J_{k|k - 1}}} {w_{k|k - 1}^{\tau ,u}q_k^{\tau ,u}\left( {\boldsymbol{z}} \right)} }},
\end{align}
and
\begin{align}
\left\{ \begin{gathered}
  {\boldsymbol{m}}_{k|k}^{i,u}\left( {\boldsymbol{z}} \right){\boldsymbol{ = m}}_{k|k - 1}^{i,u} + {\boldsymbol{K}}_k^{i,u}{\boldsymbol{\tilde v}}_k^{i,u}, \hfill \\
  {\boldsymbol{\tilde v}}_k^{i,u} = {\boldsymbol{z}} - {{\boldsymbol{H}}_k}{\boldsymbol{m}}_{k|k - 1}^{i,u}, \hfill \\
  {\boldsymbol{K}}_k^{i,u} = {\boldsymbol{P}}_{k|k - 1}^{i,u}{\boldsymbol{H}}_k^T{\left( {{\boldsymbol{S}}_k^{i,u}} \right)^{ - 1}}, \hfill \\
  {\boldsymbol{P}}_{k|k}^{i,u} = \left( {{\boldsymbol{I}} - {\boldsymbol{K}}_k^{i,u}{{\boldsymbol{H}}_k}} \right){\boldsymbol{P}}_{k|k - 1}^{i,u}, \hfill \\
  {\boldsymbol{S}}_k^{i,u} = {{\boldsymbol{H}}_k}{\boldsymbol{P}}_{k|k - 1}^{i,u}{\boldsymbol{H}}_k^T + {\boldsymbol{R}}_k^u, \hfill \\ 
\end{gathered}  \right.
\end{align}
4) Calculate the likelihood function of ${\boldsymbol{z}}$ 
\begin{align}
\Lambda _k^{i,u} = \frac{{{{\bar c}_u}}}{{{{\left( {2\pi } \right)}^{\tfrac{m}{2}}}{{\left| {{\boldsymbol{S}}_k^{i,u}} \right|}^{\tfrac{1}{2}}}}}\exp \left( {\tfrac{{{{\left( {{\boldsymbol{\tilde v}}_k^{i,u}} \right)}^T}{{\left( {{\boldsymbol{S}}_k^{i,u}} \right)}^{ - 1}}{\boldsymbol{\tilde v}}_k^{i,u}}}{{ - 2}}} \right).
\end{align}
Update the model probability $\vartheta _k^{i,u} = \Lambda _k^{i,u}/\sum\nolimits_{u = 1}^L {\Lambda _k^{i,u}} $.

5) Fuse the information of $L$ PHD filter
\begin{align}
\left\{ \begin{gathered}
  w_k^i = \sum\limits_{u = 1}^L {w_k^{i,u}\vartheta _k^{i,u}} ,{\boldsymbol{m}}_{k|k}^i = \sum\limits_{u = 1}^L {{\boldsymbol{m}}_{k|k}^{i,u}\vartheta _k^{i,u}} , \hfill \\
  {\boldsymbol{P}}_{k|k}^i = \sum\nolimits_{u = 1}^L {\vartheta _k^{i,u}\left[ {{\boldsymbol{P}}_{k|k}^{i,u} + {{\boldsymbol{e}}_{m,k|k}}{\boldsymbol{e}}_{m,k|k}^T} \right]} , \hfill \\ 
\end{gathered}  \right.
\end{align}
with ${{\boldsymbol{e}}_{m,k|k}} = {\boldsymbol{m}}_{k|k}^{i,u} - {\boldsymbol{m}}_{k|k}^i$.
The states and number can be extracted through $w_k^i$, ${\boldsymbol{m}}_{k|k}^i$, and ${\boldsymbol{P}}_{k|k}^i$.

\begin{remark}
The only difference in the inputs of different GM-PHD filters is the covariance matrices decomposed by the GMM model. By interactively fusing the results under different Gaussian models, one can obtain the fused state of multi-target. 
Specifically, when L=1, the IMF-GM-PHD algorithm degenerates into the classical GM-PHD filter, which suggests that the GM-PHD algorithm is a particular instance of our method.
\end{remark}

\section{Simulations}\label{simulation}
The effectiveness of the IMF-GM-PHD algorithm is validated through a two-dimensional MTT scenario with non-Gaussian measurement noise. In the considered scenario, the number of targets remains uncertain and fluctuates over time, and the targets detected by the sensor include clutter. A comparison is made between the performance of our method and the classical GM-PHD method, and all simulation results are the average outcomes of 200 independent experiments.

\begin{figure}[ht]
	\centering 
	\includegraphics[width=1.0\linewidth]{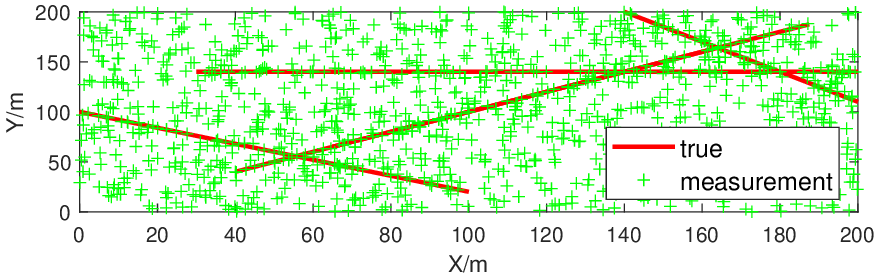} 
	\caption{Measurement data and true trajectories} 
	\label{t_measurement} 
\end{figure}
\begin{figure}[ht]
	\centering 
	\includegraphics[width=1.0\linewidth]{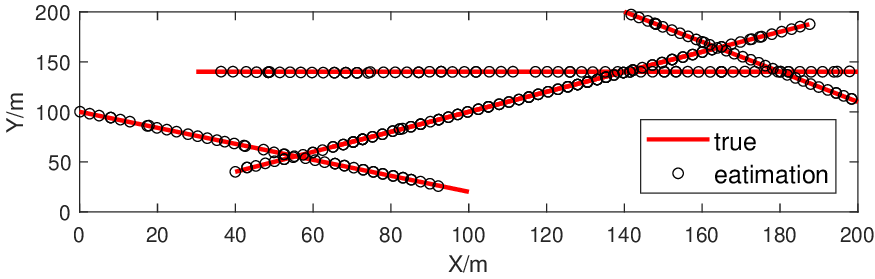} 
	\caption{Position estimates of the proposed algorithm} 
	\label{resultp} 
\end{figure}

The tracked targets \cite{10239447,huang2023distributed} are modeled as:
\begin{align}
\left\{ \begin{gathered}
  {{\boldsymbol{x}}_k} = {{\boldsymbol{F}}_{k|k - 1}}{{\boldsymbol{x}}_{k - 1}} + {{\boldsymbol{q}}_{k - 1}}, \hfill \\
  {{\boldsymbol{z}}_k} = {{\boldsymbol{H}}_k}{{\boldsymbol{x}}_k} + {{\boldsymbol{r}}_k}, \hfill \\ 
\end{gathered}  \right.
\end{align}
where ${{\boldsymbol{F}}_{k|k - 1}} = \left[ {\begin{array}{*{20}{c}}
  1&{\Delta t}&0&0 \\ 
  0&1&0&0 \\ 
  0&0&1&{\Delta t} \\ 
  0&0&0&1 
\end{array}} \right]$ is the state transition matrix, ${{\boldsymbol{H}}_k} = \left[ {\begin{array}{*{20}{c}}
  {\begin{array}{*{20}{c}}
  1&0&0&0 
\end{array}} \\ 
  {\begin{array}{*{20}{c}}
  0&0&1&0 
\end{array}} 
\end{array}} \right]$ is measurement matrix. ${{\boldsymbol{x}}_k} = {\left[ {\begin{array}{*{20}{c}}
  {{x_{k,1}}}&{{x_{k,2}}}&{{x_{k,3}}}&{{x_{k4}}} 
\end{array}} \right]^T}$ represents the state of each target, ${{x_{k,1}}}$ and ${{x_{k,3}}}$ are the position on the $x$ and $y$ axes, ${{x_{k,2}}}$ and ${{x_{k4}}}$ are the velocity on the $x$ and $y$ axes. The time interval is $\Delta t = 1$ second, ${p_{S,k}} = 0.99$, and ${p_{D,k}} = 0.98$. Clutter conforms to a Poisson distribution, and its average density is 10 per unit volume across the region within $\left[ {0{\text{ 200}}} \right] \times \left[ {0{\text{ 200}}} \right]$. The process noise follows a Gaussian distribution, and its mean and covariance is zero and ${{\boldsymbol{Q}}_k} = diag\left[ {\begin{array}{*{20}{c}}
  {0.01}&{0.1}&{0.01}&{0.1} 
\end{array}} \right]$. Measurement noise is set to a classical non-Gaussian noise with heavy-tail distribution \cite{he2023generalized11} $0.7\mathcal{N}\left( {0,0.01} \right) + 0.3\mathcal{N}\left( {0,100} \right)$.

\begin{figure}[ht]
	\centering 
	\includegraphics[width=1.0\linewidth]{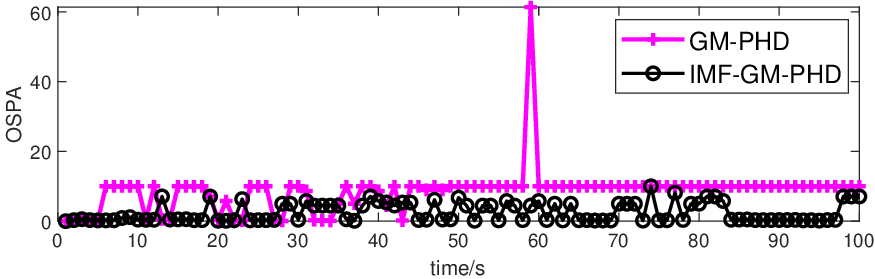} 
	\caption{The OSPA of the GM-PHD and IMF-GM-PHD filters.} 
	\label{ospa} 
\end{figure}

Fig. \ref{t_measurement} displays the genuine paths of four targets, accompanied by sensor data containing clutter. Moreover, the measurement data originating from real targets is also contaminated by non-Gaussian noise. Fig. \ref{resultp} illustrates the results of the IMF-GM-PHD filter for multi-target states. In this study, the performance of both the IMF-GM-PHD and the GM-PHD filters is assessed utilizing the optimal sub-pattern assignment (OSPA) metric \cite{4567674}. The numerical simulation results indicate that 1) the IMF-GM-PHD filter presents satisfactory estimates for multi-target states; 2) the OSPA of our method is notably lower than that of the GM-PHD filter, which means that IMF-GM-PHD filter performs better than the classical GM-PHD filter with non-Gaussian noise.

\section{Conclusion}
\label{conclusion}
In this study, an interactive model fusion-based GM-PHD filter is designed. The key idea of the IMF-GM-PHD filter is to transform the MTT problem under non-Gaussian conditions into an MTT problem under Gaussian conditions. The conventional GM-PHD filter serves as a fundamental component of the proposed algorithm, and the concept of interactive multiple models is adopted to achieve the fusion between different components. Simulation validates the feasibility and effectiveness of the IMF-GM-PHD filter under non-Gaussian noise. 
\vfill\pagebreak

\bibliographystyle{IEEEbib}
\bibliography{refs}
\end{document}